\documentstyle{article}

\def\ba{\begin{array}}
\def\ea{\end{array}}
\def\be{\begin{equation}}
\def\ee{\end{equation}}

\def\D{{\Delta }}
\def\P{{\Phi}}

\def\p{{\phi}}
\def\a{{\alpha}}

\def\d{{\rm d}}

\def\bea{\begin{eqnarray}}
\def\eea{\end{eqnarray}}

\def\unit{\rm {1\kern-.4em 1}}

\begin{document}
\begin{titlepage}
\vspace{-10mm}
\vspace{12pt}
\hfill{}
\vskip 5 mm
\leftline{ \Large \bf
          Logarithmic Conformal Field Theories with Continuous Weights}
\vskip 1.5 cm
\leftline{\bf M. Khorrami$^{1,2,3}$, A. Aghamohammadi$^{1,4}$, 
              M. R. Rahimi Tabar$^{1,5,*}$}
\vskip 5 mm
{\it
  \leftline{ $^1$ Institute for Studies in Theoretical Physics and
            Mathematics, P.O.Box  5531, Tehran 19395, Iran. }
  \leftline{ $^2$ Department of Physics, Tehran University,
             North Kargar Ave. Tehran, Iran. }
  \leftline{ $^3$ Institute for Advanced Studies in Basic Sciences,
             P.O.Box 159, Gava Zang, Zanjan 45195, Iran. }
  \leftline{ $^4$ Department of Physics, Alzahra University,
             Tehran 19834, Iran. }
 \leftline{ $^5$ Department of Physics, University of Science and
            Technology, Narmak, Tehran 16844, Iran}
  \leftline{ $^*$ Rahimi@netware2.ipm.ac.ir}
  }
\vskip 5 cm
\begin{abstract}
We study the  logarithmic conformal field theories in which conformal weights 
are continuous subset of real numbers. A general relation between the 
correlators consisting of logarithmic fields and those consisting of ordinary
conformal fields is investigated. As an example the correlators of the 
Coulomb-gas model are explicitly studied.
\end{abstract}
\vskip 10 mm
\end{titlepage}
{\section {Introduction}}
It has been shown by Gurarie \cite{Gu}, that conformal field theories
(CFTs) whose correlation functions exhibit logarithmic behaviour, can be
consistently defined and if in the OPE of two given local
fields which has at least two fields with the same conformal dimension, one
may find some operators with a special property, known as logarithmic
operators. As discussed in \cite{Gu}, these operators with the ordinary
operators form the basis of the Jordan cell for the operators $L_i$.

The logarithmic fields (operators) in CFT were first studied by Gurarie 
in the $c=-2$ model \cite{Gu}. After Gurarie, thes logarithms have been 
found in a multitude of others models such as the WZNW--model on 
GL(1,1) [2], the gravititionally dressed CFT s [3], $c_{p,1}$ and non--minimal
$c_{p,q}$ models [2, 4--6], critical disorderd models [7,8], and the 
WZNW--models
at level 0 [9,10]. They play a role in the study of critical polymers and 
percolation [11,12], 2D--MHD turbulence [13--15], 2D--turbulence [16,17]
and quantum Hall states [18--20]. They are also important for studing the 
problem of recoil in the string theory and D--branes [9, 21--24], 
as well as target space symmetries in string theory [9].  
The representation theory of the Virasoro algebra for LCFT s was developed 
in [25]. The origin of the LCFT s has been discussed in [26--28].
The modular invariant partition functions for
$c_{eff}=1$ and the fusion rules of logarithmic conformal field
theories (LCFT) are considered in [4], see also [29] about 
consequences for Zamolodchikov's C--theorem. Structure of the LCFT s in 
D--dimensions has been discussed in [30].

The basic properties of logarithmic operators are that,
they form a part of the basis of the Jordan cell
for $L_i$'s and in the correlator of such fields there
is a logarithmic singularity \cite{Gu}.
It has been shown that in rational minimal models such a situation,
i.e. two fields with the same dimensions, doesn't occur [14].

In a previous paper 
\cite{RAK} assuming conformal invariance we have explicitly calculated two- 
and three-point functions for the case of more than one logarithmic field 
in a block, and more than one set of logarithmic fields for the case where 
conformal weights belong to a discrete set. Regarding logarithmic fields 
{\it formally} as derivations of ordinary fields with respect to their 
conformal dimension, we have calculated n-point functions containing 
logarithmic fields in terms of those of ordinary fields (see also [31],
about the role of such derivative in the OPE coefficients of LCFT).

We have done these when conformal weights 
belong to a discrete set. 
In [28], there 
is an  attempt to understand the meaning of derivation CFT's with respect to 
conformal weights.
Here, we want to consider logarithmic conformal field theories with continuous 
weights. The simplest example of such theories  is the free field theory. 
The structure of this article is as follows. In section {\bf 2} 
we study conformal theories, in which conformal weights belong to a continuous 
subset of real numbers, and calculate the correlators of these theories. 
Specifically, we show that one can calculate the two-point functions of 
logarithmic fields in terms of those of ordinary fields by derivation. This 
is not possible in the case of discrete weights. In section {\bf 3} we 
consider the Coulomb-gas model as an example. 

{\section {Correlators of a logarithmic CFT with continuous weights}}

In \cite{RAK}, it was shown that if there are quasi-primary fields in a 
conformal field theory, there arises logarithmic terms in the correlators of  
the theory. By quasi-primary fields, it is meant a family of operators 
satisfying

\be \label{1}
[L_n,\P^{(j)}(z)]=z^{n+1}\partial_z\P^{(j)}(z)+(n+1)z^n\D \P^{(j)}(z)+
(n+1)z^n\D \P^{(j-1)}(z),
\ee
where $\D$ is the conformal weight of the family. Among the fields $\P^{(j)}$,
the field $\P^{(0)}$ is primary.
It was shown that one can interpret the fields $ \P^{(j)}$, {\it formally}, 
as the $j$-th derivative of a field with respect to the conformal weight:
\be 
\P^{(j)}(z)={1\over j!}{\d ^j\over \d \D^j}\P^{(0)}(z),
\ee
and use this to calculate the correlators containing $\P^{(j)}$ in terms of 
those containing $\P^{(0)}$ only. The transformation relation (\ref{1}), 
and the symmetry of the theory under the transformations generated by 
$L_{\pm 1}$ and $L_0$, were also exploited to obtain two-point functions 
for the case where conformal weights belong to a discrete set. 
There were two features in two point functions. First, for two families 
$\P_1$ and $\P_2$, consisting of $n_1+1$ and
$n_2+1$ members, respectively, it was shown that the correlator 
$<\P^{(i)}_1\P^{(j)}_2>$ is zero unless $i+j\geq {\rm max} (n_1, n_2)$.
(It is understood that the conformal weights of these two families are equal.
Otherwise, the above correlators are zero.) Another point was that 
one could not use the derivation process with respect to the conformal 
weights to obtain the two-point functions of these families from 
$<\P_1^{(0)}\P_2^{(0)}>$, since the correlators contain a multiplicative term 
$\delta_ {\D_1, \D_2}$, which can not be differentiated with respect to the 
conformal weight.

Now, suppose that the set of conformal weights of the theory is a continous 
subset of the real numbers. First, reconsider the arguments resulted to the 
fact that $<\P_1^{(i)}\P_2^{(j)}>$ is equal to zero for $i+j\geq 
{\rm max} (n_1, n_2)$. These came from the symmetry of the theory under 
the action of $L_{\pm 1}$ and $L_0$. Symmetry under the action of $L_{-1}$ 
resutlts in 
\be
<\P_1^{(i)}(z)\P_2^{(j)}(w)>=<\P_1^{(i)}(z-w)\P_2^{(j)}(0)>=:A^{ij}(z-w).
\ee
We also have 
\be \label{4}
<[L_0,\P_1^{(i)}(z)\P_2^{(j)}(0)]>=(z\partial +\D_1 +\D_2)A^{ij}(z) + 
                                   A^{i-1,j}(z) +A^{i,j-1}(z)=0,
\ee
and
\be
<[L_1,\P_1^{(i)}(z)\P_2^{(j)}(0)]>=(z^2\partial +2z\D_1 )A^{ij}(z) + 
                                   2zA^{i-1,j}(z) =0.
\ee
These show that 
\be \label{6}
(\D_1 -\D_2)A^{ij}(z) + A^{i-1,j}(z) -A^{i,j-1}(z)=0.
\ee                                                
If $\D_1 \ne \D_2$, it is easily seen, through a recursive calculation, that 
$A^{ij}$'s are all equal to zero.
This shows that the support of these correlators, as distribution  of $\D_1$ 
and $\D_2$, is $\D_1 -\D_2=0$.
So, one can use the ansatz
\be \label{7}
A^{ij}(z)=\sum _{k\geq 0} A^{ij}_k(z)\delta^{(k)}(\D_1 -\D_2).
\ee
Inserting this in (\ref{6}), and using $x\delta^{(k+1)}(x)=-(k+1)
\delta^{(k)}(x)$, it is seen that 
\be \label{8}
\sum_{k\geq 0}[-(k+1)A^{ij}_{k+1}(z) + A^{i-1,j}_k(z) -A^{i,j-1}_k(z)]
\delta^{(k)}(\D_1 -\D_2)=0,
\ee                                                
or
\be \label{9}
(k+1)A^{ij}_{k+1}(z)= A^{i-1,j}_k(z) -A^{i,j-1}_k(z),  \qquad k\geq 0
\ee                                                
This equation is readily solved:
\be \label{10}
A^{ij}_k(z)={1\over k!}\sum_{l=0}^k (^k_l) A^{i-k+l,j-l}_0(z),
\ee
where $ A^{ij}_0$'s remain arbitrary. Also note that $A^{ij}_k$'s with a 
negative index are zero. We now put (\ref{7}) in (\ref{4}). This gives 
\be \label{11}
(z\partial +\D_1 +\D_2)A^{ij}_k(z) + A^{i-1,j}_k(z) +A^{i,j-1}_k(z)=0,
\ee
Using (\ref{10}), it is readily seen that it is sufficient to write 
(\ref{11})  only for $k=0$ . 
This gives 
\be \label{12}
(z\partial +\D_1 +\D_2)A^{ij}_0(z) + A^{i-1,j}_0(z) +A^{i,j-1}_0(z)=0.
\ee
Putting the ansatz
\be
A^{ij}_0(z)=z^{-(\D_1+\D_2)}\sum ^{i+j}_{m=0}\a^{ij}_m (\ln z)^m
\ee
in (\ref{12}), one arrives at 
\be \label{14}
(m+1)\a^{ij}_{m+1} + \a^{i-1,j}_m +\a^{i,j-1}_m=0,
\ee
the solution to which is
\be
\a^{ij}_m={(-1)^m\over m!}\sum^m_{s=0}(^m_s)\a^{i-m+s,j-s}_0.
\ee
From this 
\be
A^{ij}_0(z)=z^{-(\D_1+\D_2)}\sum ^{i+j}_{m=0} (\ln z)^m 
{(-1)^m\over m!}\sum^m_{s=0}(^m_s)\a^{i-m+s,j-s}_0,
\ee
and
\be
A^{ij}_k(z)=[{1\over k!}\sum ^k_{l=0}(-1)^l (^k_l)
\sum ^{i+j-k}_{m=0} (\ln z)^m 
{(-1)^m\over m!}\sum^m_{s=0}(^m_s)\a^{i-k-m+l+s,j-l-s}_0]z^{-(\D_1+\D_2)}.
\ee
So we have
\be 
A^{ij}(z)=z^{-(\D_1+\D_2)}\sum _{k\geq 0} \delta^{(k)}(\D_1 -\D_2)
[{1\over k!}\sum ^k_{l=0}(-1)^l (^k_l)
\sum ^{i+j-k}_{m=0} (\ln z)^m 
{(-1)^m\over m!}\sum^m_{s=0}(^m_s)\a^{i-k-m+l+s,j-l-s}_0],
\ee
or 
\be \label{19}
A^{ij}(z)=z^{-(\D_1+\D_2)}\sum _{p,q,r,s\geq 0} 
{(-1)^{q+r+s}\over p!q!r!s!}\a^{i-p-r,j-q-s} (\ln z)^{r+s}
\delta^{(p+q)}(\D_1 -\D_2),
\ee
where
\be
\a^{ij}:=\a^{ij}_0.
\ee
These constants, defined for nonnegative values of $i$ and $j$, are arbitrary
and not determined from the conformal invariance only.

Now differentiate (\ref{19}) formally with respect to $\D_1$. In this process,
$\a^{ij}$'s are also assumed to be functions of $\D_1$ and $\D_2$. 
This leads to 
\be \ba{ll}\label{21}
{\partial A^{ij}(z)\over \partial \D_1}=&z^{-(\D_1+\D_2)}\sum _{p,q,r,s} 
{(-1)^{q+r+s}\over p!q!r!s!}\{ {\partial \a^{i-p-r,j-q-s}\over \partial \D_1}
(\ln z)^{r+s}\delta^{(p+q)}(\D_1 -\D_2)
+\a^{i-p-r,j-q-s}\cr & [(\ln z)^{r+s}\delta^{(p+q+1)}(\D_1 -\D_2)-
(\ln z)^{r+s+1}\delta^{(p+q)}(\D_1 -\D_2)]\},
\ea 
\ee
or 
\be 
{\partial A^{ij}(z)\over \partial \D_1}=z^{-(\D_1+\D_2)}\sum _{p,q,r,s} 
{(-1)^{q+r+s}\over p!q!r!s!}(\ln z)^{r+s}\delta^{(p+q)}(\D_1 -\D_2)
[(p+r) \a^{i-p-r,j-q-s}+{\partial \a^{i-p-r,j-q-s} \over \partial \D_1}]. 
\ee
Comparing this with $A^{i+1,j}$, it is easily seen that 
\be
A^{i+1,j}={1\over i+1}{\partial A^{ij}\over \partial \D_1},
\ee 
provided
\be  \label{24}
{\partial \a^{i-p-r,j-q-s} \over \partial \D_1}=(i+1-p-r)\a^{i+1-p-r,j-q-s}  .
\ee
Note, however, that the left hand side of (\ref{24}) is just a {\it formal 
differentiation}. That is, the functional dependence of $\a^{ij}$'s on 
$\D_1$ and $\D_2$ is not known, and their derivative is just another 
constant. Repeating this procedure for $\D_2$, we finally arrive at 
\be
\a^{ij}={1\over i!j!}{\partial^i \over \partial \D_1^i}
{\partial^j \over \partial \D_2^j}\a^{00},
\ee
and 
\be
A^{ij}={1\over i!j!}{\partial^i \over \partial \D_1^i}
{\partial^j \over \partial \D_2^j}A^{00}.
\ee
These relations mean that one can start from $A^{00}$, which is simply 
\be 
 A^{00}(z)=z^{-(\D_1+\D_2)}\delta (\D_1 -\D_2)
 \a^{00}, 
\ee
and differentiate it with respect to $\D_1 $ and $\D_2$, to obtain $A^{ij}$.
In each differentiation, some new constants appear, which are denoted by 
$\a^{ij}$'s but with higher indices. Note also that the definition is 
self-consistent. So that this formal differentiation process is well-defined. 

One can use this two-point functions to calculate the one-point functions of 
the theory. We simply put $\P^{(0)}_2=1$. So, $\D_2=0$,
\be
<\P^{(0)}(z)>=\beta^0\delta (\D),
\ee
and
\be
<\P^{(i)}(z)>=\sum^i_{k=0}{\beta^{n-k}\over k!}\delta^k(\D),
\ee
where
\be
\beta^i:={1\over i!}{\d^i \beta^0\over \d \D^i}.
\ee
The more than two-point function are calculated exactly the same as in 
\cite{RAK}.
\

\

\

{\section { The Coulomb--gas model as an example of LCFT's}}
\noindent As an explicit example of the general formulation of the previous 
section,
consider the Coulomb-gas model characterized by the action [26]
\be \label{31}
S={1\over 4\pi} \int \d^2 x \sqrt g [-g^{\mu \nu }(\partial_{\mu}  \P )
(\partial_{\nu} \P ) +i QR\P ],
\ee
where $\P$ is a real scalar field,  $Q$ is the charge of the theory,
$R$ is the scalar curvature of the surface and the surface itself is of 
a spherical topology, and is everywhere flat except at a single point. 

Defining the stress tensor as 
\be
T^{\mu \nu}:=-{4\pi \over \sqrt g}{\delta S\over \delta g_{\mu \nu}},
\ee
it is redily seen that 
\be 
T^{\mu \nu}=-(\partial^{\mu}  \P )(\partial^{\nu} \P ) 
+{1\over 2} g^{\mu \nu} g^{\a \beta}(\partial_{\a}  \P )(\partial_{\beta} 
\P )-iQ [\p^{;\mu \nu }-g^{\mu \nu}\nabla^2\P ],
\ee
and
\be \label{34}
T(z):=T_{zz}(z)=-(\partial \p )^2-iQ\partial^2 \p ,
\ee
where in the last relation the equation of motion has been used to write 
\be
\P (z, {\bar z})=\p (z)+{\bar \p }({\bar z}).
\ee
It is well known that this theory is conformal, with the central charge
\be 
c=1-6Q^2
\ee
There are, however, some features which need more care in our later 
calculations. First, this theory can not be normalized so that the 
expectation value of the unit operator become unity.
In fact, using $e^S$ as the integration measure, it is seen that 
\be 
<1>\propto \delta (Q)
\ee
one can, at most, normalize this so that 
\be \label{38}
<1>= \delta (Q)
\ee
Second, $\p$ has a $z$-independent part, which we denote it by $\p_0$.
The expectation value of $\p_0$ is not zero. In fact, from the action 
(\ref{31}), 
\be
<\p >=<\p_0 >={1\over N(Q)}\int \d \p_0 \p_0 \exp (2iQ\p_0),
\ee
where $N$ is determined from (\ref{38}) and 
\be
<1>={1\over N}\int \d \p_0 \exp (2iQ\p_0).
\ee
This shows that $N(0) =\pi $, and 
\be 
<\p_0 >={1\over 2i}[\delta ' (Q) +{N'(0)\over N(0)}\delta (Q)]
\ee
More generally
\be
<f(\p_0 )>={1\over N}f({1\over 2i}{\d \over \d Q} ) (N<1>)
          ={1\over N}f({1\over 2i}{\d \over \d Q} ) [N\delta (Q)].
\ee

Third, the normal ordering procedure is defined as following.
One can write 
\be
\p (z)= \p_0+\p_+ (z) + \p_- (z),
\ee
where $<0|\p_-(z)=0, \ \ \p_+(z)|0>=0$, and 
\be 
[\p_0,\p_{\pm}]=0.
\ee
The normal ordering is so that one puts all `-' parts at the left of all 
`+' parts. It is then seen that 
\be \label{46}
<:f[\p ]:>=<f(\p_0 )>
\ee
Here, the dependence of $f$ on $\p $ in the left hand side may be quite 
complicated; even $f$ can depend on the values of $\p $ at different points.
In the right hand side, however, one simply changes $\p (z) \to \p_0$.

Now consider the two point funcion. From the equation of motion, we have 
\be \label{47}
<\p (z)\p (w)>=-{1\over 2}\ln (z-w) <1> +b
\ee
we also have 
\be
<:\p (z)\p (w):>=<\p_0^2>= -{1\over 4N} {\d^2 \over \d Q^2 }[ N\delta (Q)]
\ee
Note that there is an arbitrary term in (\ref{47}), due to the ultraviolet
divergence of the theory. 
One can use this arbitrariness, combined with the arbitrariness in $N(Q)$,
to redefine the theory as 
\be \label{48}
\p (z)\p (w)=:-{1\over 2}\ln (z-w) +:\p (z)\p (w):,
\ee
and 
\be 
<f(\p_0)>:=f({1\over 2i} {\d \over \d Q})\delta (Q)
\ee
these relations, combined with (\ref{46}) are sufficient 
to obtain all of the correlators. One can, in addition, use (\ref{34})
(in normal ordered form) to arrive at 
\be \label{51}
T(z) \p (w)={\partial _w \p\over z-w}-{iQ/2\over (z-w)^2}+{\rm r.t.},
\ee
and 
\be 
T(z) T(w)={\partial _w T\over z-w}-{2T(w)\over (z-w)^2}+
{(1-6Q^2)/2\over (z-w)^4}.
\ee
Eq. (\ref{51}) can be written in the form 
\be
[L_n,\p (z)]=z^{n+1}\partial \p -{iQ\over 2}(n+1)z^n.
\ee
This shows that the operators $\p $ and $1$ are a pair of logarithmic 
operators with $\D =0$ (in the sence of (\ref{1})). One can easily show that 
\be \label{54}
T(z) :{\rm e}^{i\a \p (w)}: \ ={\partial _w :{\rm e}^{i\a \p (w)}:\over z-w}-
{\a (\a +2Q)/4\over (z-w)^2}:{\rm e}^{i\a \p (w)}: + {\rm r.t.},
\ee
which shows that $:{\rm e}^{i\a \p }:$ is a primary field with 
\be 
\D_{\a}={\a (\a +2Q) \over 4}
\ee
To this field, however, there corresponds a quasi conformal family 
(pre--logarithimic operators [26]), whose 
members are obtained by explicit differentiation with respect to $\a$ 
($\a$ is not the conformal weight but is a function of it):
\be
W_{\a }^{(n)}:\ =\ :\p^n {\rm e}^{i\a \p }:\ = (-i)^n {\d \over \d \a^n }
:{\rm e}^{i\a \p }:.
\ee

To calculate the correlators of $W$'s, it is sufficient to calculate                       
$<W_{\a_1}^{(0)}\cdots W_{\a_k}^{(0)}>$.

One has, using Wick's theorem and (\ref{48}),
\be
\Pi^k_{j=1} :{\rm e}^{i\a_j \p (z_j)}: \ ={\rm e}^{1/2 \sum_{1\leq i<j\leq k}
\a_i \a_j\ln (z_i-z_j)}:{\rm e}^{i\sum_{j=1}^k\a_j \p (z_j)}:
\ee
From  this using (\ref{46}) and (\ref{48}), we have
\be \ba{ll}
<\Pi_{j=1}^kW_{\a_j}^{(0)}(z_j)>&=
[\Pi_{ \scriptstyle{1\leq i<j\leq k}}(z_i-z_j)^{\a_i \a_j\over 2}]
{\rm e}^{1/2\sum_{j=1}^k\a_j{\d \over \d Q} }\delta (Q)\cr
&=[\Pi_{\scriptstyle{1\leq i<j\leq k}}(z_i-z_j)^{\a_i \a_j\over 2}]
\delta (Q+{1\over 2} \sum_{j=1}^k \a_j).
\ea 
\ee
Obviously, differentiating with respect to any $\a_i$, leads to logarithmic 
terms for  the correlators consisting of logarithmic fields $W_{\a}^{(n)}$.
The power of logarithmic terms is equal to the sum of superscripts of the 
fields $W_{\a}^{(n)}$.
                      
\newpage

\end{document}